# Fabrication and spectral characterization of the porous dielectric THz waveguides using microstructured molding technique.

A. Dupuis, A. Mazhorova, F. Desevedavy, M. Rosé, M. Skorobogatiy
Ecole Polytechnique de Montréal, Montréal, Québec, Canada
www.photonics.phys.polymtl.ca

**ABSTRACT**

We report two novel fabrication techniques, as well as spectral transmission and propagation loss measurements of the subwavelength plastic wires with highly porous (up to 86%) and non-porous transverse geometries. The two fabrication techniques we describe are based on the microstructured molding approach. In one technique the mold is made completely from silica by stacking and fusing silica capillaries to the bottom of a silica ampoule. The melted material is then poured into the silica mold to cast the microstructured preform. Another approach uses microstructured mold made of plastic which is co-drawn with a cast preform. Material of the mold is then dissolved after fiber drawing. We also describe a novel THz-TDS setup with an easily adjustable optical path length, designed to perform cutback measurements using THz fibers of up to 50 cm in length. We find that while both porous and non-porous subwavelength fibers of the same outside diameter have low propagation losses ($\alpha \leq 0.02 \text{cm}^{-1}$), however, the porous fibers exhibit a much wider spectral transmission window and enable transmission at higher frequencies compared to the non-porous fibers. We then show that the typical bell-shaped transmission spectra of the subwavelengths fibers can be very well explained by the onset of material absorption loss at higher frequencies due to strong confinement of the modal fields in the material region of the fiber, as well as strong coupling loss at lower frequencies due to mismatch of the modal field diameter and a size of the gaussian-like beam of a THz source.

## 1. INTRODUCTION

In order to enable various stand-off and remote THz applications[1] a considerable amount of research has being focused on developing low-loss waveguides and fibers as a key technology for the flexible and convenient THz light delivery. Given relatively high material losses of the dielectric and metallic materials in THz region it seems that the only viable solution for designing low-loss THz waveguides is through introduction of the low-loss gas-filled hollow regions into the waveguide structure, and maximization of the portion of guided power in such regions. The simplest waveguide that operates on this principle is a subwavelength wire featuring large evanescent fields that extends far beyond the waveguide core and into the surrounding gaseous cladding. Both the metallic[2] and dielectric[3] subwavelength fibers have been demonstrated to have propagation losses on the order of 0.01 cm$^{-1}$ in the vicinity of 0.3 THz, which are among the lowest losses reported to date. Metallic wires, in particular, have been shown to support Sommerfeld plasmons[2,4,5] offering a low dispersion, low propagation loss single-mode regime over a wide range of frequencies. However, the azimuthally polarized Sommerfeld mode is difficult to excite and specialized sources are required to increase the coupling efficiency.[6] Recent efforts on metallic wires have demonstrated the use of a dual-wire suspended in air waveguide,[8] which supports a more easily excitable TEM mode at the expense of a more complicated waveguide geometry that requires maintaining a constant sub-millimeter distance between the two wires. On the other hand, dielectric subwavelength wires[3,9–11] feature low-loss single-mode regime with a linearly-polarized HE$_{11}$ fundamental mode which is very easy to excite with a gaussian-like beam of a THz source. Main disadvantage of the dielectric wires is a relatively high dispersion, and a limited operational bandwidth due to onset of large absorption losses at higher frequencies. Moreover, both types of subwavelength fiber feature extremely delocalized modes (as large as several mm in diameter), which are susceptible to high bending loss and strong proximity cross-talk with the environment. In some implementations, strong mode delocalization can be used for a benefit to enable, for example, non-destructive cut-back-like measurements of the fiber propagating losses using evanescent directional coupler technique[10,12]. Another potential benefit of the subwavelength waveguides is in THz optical sensing which rely on strong presence of the modal evanescent fields in the analyte.[13] While bending loss of the subwavelength fibers has been shown to be very high[5], the deflection loss[7,11] (micro bending at a single point point) could be relatively small provided that the deflection angle is also small (<2°). The group of Sun et al. has exploited low-loss fiber deflections to make a THz fiber-based imaging[10, 11] and a near-field microscope.[14]

In this paper, we report several fabrication techniques, as well as spectral transmission and propagation loss measurements of the subwavelength plastic wires with various porous and non-porous transverse geometries. Our original theoretical investigations[15–16] have predicted that addition of an array of subwavelength holes into the crossection of a subwavelength fiber not only increases the fraction of power guided in the air (thus leading to a lower absorption loss), but also allows to concentrate more light within the porous fiber core, thus, dramatically reducing the proximity cross-talk of the fiber mode with the environment. Moreover, we have demonstrated that porous fibers exhibit a much smaller bending loss compared to the non-porous fibers of the comparable optical properties, which was explained by stronger modal confinement in the porous fibers. Finally, we have demonstrated that porous fibers with outer diameters much larger than the wavelength of light can be designed, while still operating in a single mode regime and exhibiting very low propagation and bending loss[16]. To design such fibers one has to use very high porosity and strongly subwavelength material veins. Our findings were later reproduced by another group[17]. One of the practical benefits for the use of porous fibers compared to the non-porous fibers of the same propagation loss is the much large size of porous fibers[16], which greatly simplifies their handling. We have recently published experimental results where porous fibers were fabricated and their spectrally averaged transmission losses were characterized using novel Directional Coupler Method.[12] In that work, however, spectral information was inferred indirectly as we have used a broadband THz source and a balometer detector to perform cut-back-like measurements of the fiber losses. Moreover, our original fiber fabrication strategies resulted in fibers with relatively low porosities below 40%.

In this paper, we present comparative study of the spectrally resolved loss measurements of porous and non-porous subwavelength fibers performed using a novel THz-TDS (Time Domain Spectroscopy) setup of adjustable optical path length specifically designed for the fiber measurements. Moreover, we report a novel fabrication method using microstructured mold casting which allowed us to fabricate fibers of very high 86% porosity. Although both porous and nonporous fibers of the same diameter show very low propagation losses below $0.02cm^{-1}$, we find, however, that the porous fibers exhibit a much wider spectral transmission window and enable transmission at higher frequencies compared to the non-porous fibers. We then show that the typical bell-shaped transmission spectra of the subwavelengths fibers can be very well explained by the onset of material absorption loss at higher frequencies due to strong confinement of the modal fields in the material region of the fiber, as well as onset of a strong coupling loss at lower frequencies due to mismatch of the modal field diameter and a size of the gaussian-like beam of a THz source.

The paper is organized as follows. Section 2 presents a variable optical path length THz-TDS setup that can accommodate fibers up to 50 cm in length. Section 3 details principles of operation of a porous subwavelength fiber, and description of the two methods for the fabrication of such fibers. Section 4 presents spectral transmission and loss measurements. Section 5 develops theoretical justification of the observed transmission spectra and measured losses. The final section concludes with a discussion on fabrication difficulties and the effect of fiber imperfections on the transmission through subwavelength fibers.

## 2. Measurement Setup

The standard THz-TDS setup, using off-axis parabolic mirrors in a focal-point-to-focal-point configuration, was developed to measure both small point samples and extended fiber samples. In order to conveniently measure the transmission spectra of THz waveguides, an easily reconfigurable setup was needed in order to accommodate waveguides of different length. Schematic of a solution that was developed in our group is presented in Figure 1(a). It features two sets of rails with mirror assemblies to allow convenient adjustment of a THz optical path (rail 2), and to allow insertion of a THz fiber (rail 1). A fixed parabolic mirror focuses the THz radiation into the waveguide, therefore in-coupling plane is fixed. The output of a waveguide is placed in the focal point of another parabolic mirror which can be displaced along the rail 1 together with a flat mirror. Light collected and collimated by the parabolic mirror is then redirected towards the fixed detector with a flat mirror. In our setup, waveguides of length up to 50 cm can be measured by simply translating the position of these two mirrors along the rail 1. Two low temperature grown GaAs photo-conductive antennae from Menlo Systems GmbH were used as both emitter and detector. These were pumped by a frequency-doubled C fiber laser, also from Menlo Systems GmbH. The emitter pump beam was chopped and the detected signal was measured with a Stanford Research Systems lock-in amplifier. The various waveguides studied were held in place with 3-axis positioning mounts, and the entire assembly was housed in a nitrogen purged cage to reduce the effects of water vapor. Figure 1(b) presents the spectrum of the source as well as the background noise level. The system had an amplitude dynamic range of more than 20 dB. Although much effort was expended trying to properly purge the cage with dry nitrogen, some residual water vapor absorption lines were still detected in the spectrum (see black lines for reference).

Note that rail 2 is generally necessary as an additional delay line to compensate for the change in the optical path introduced by the waveguide. Rail 2 assembly is especially important when measuring long waveguides that support modes of refractive indices which are significantly different from 1.0. In this case the path lengths of THz light in an empty setup (reference) and in a setup with fiber could be so different that the maximal pulse delay offered by a computer controlled variable delay line could become insufficient when trying to measure both an empty setup and a setup with fiber in it. Interestingly, in the case of subwavelength fibers the refractive index of a guided mode is very close to 1.0. Therefore, additional pulse delay introduced by the waveguide (as compared to an empty setup) is typically easily compensated by the variable delay line. Therefore, in all the measurements reported in this paper the mirrors on rail 2 were fixed, and therefore, the physical length between a source and a detector (denoted as $L_{path2}$) was fixed. In this case waveguide transmission T can be described as:

$$E_{waveguide}(\omega) = E_{source}(\omega) \cdot \eta \cdot C_{in} \cdot C_{out} \cdot e^{-i[n_{eff}(\frac{\omega}{c})L_w + \frac{\omega}{c}(L_{path2}-L_w)]} e^{-\frac{\alpha L_w}{2}}, \qquad (1)$$

$$E_{reference}(\omega) = E_{source}(\omega) \cdot \eta \cdot e^{-i\frac{\omega}{c}L_{path2}}, \qquad (2)$$

$$T = \frac{E_{waveguide}(\omega)}{E_{reference}(\omega)} = C_{in} \cdot C_{out} \cdot e^{-i(n_{eff}-1)(\frac{\omega}{c})L_w} e^{-\frac{\alpha L_w}{2}}, \qquad (3)$$

where Ewaveguide(ω) is the modal field coming out of a waveguide of length Lw, and Ereference(ω) is the reference signal measured in the absence of the waveguide at the position of a coupling plane. Esource(ω) is the spectrum of the source, whereas η is the transfer function describing transmission through our setup in the absence of a waveguide. Cin and Cout are respectively the input and output coupling coefficients with respect to the waveguide, while α and neff are the power propagation loss and effective refractive index of the mode guided by the waveguide. Note that transmission through the waveguide induces a time delay of $(n_{eff}-1)L_w/c$ with respect to a reference THz pulse propagating through an empty setup. This delay is small in the case of subwavelength fibers as for the modes of such a waveguide $n_{eff} \sim 1$.

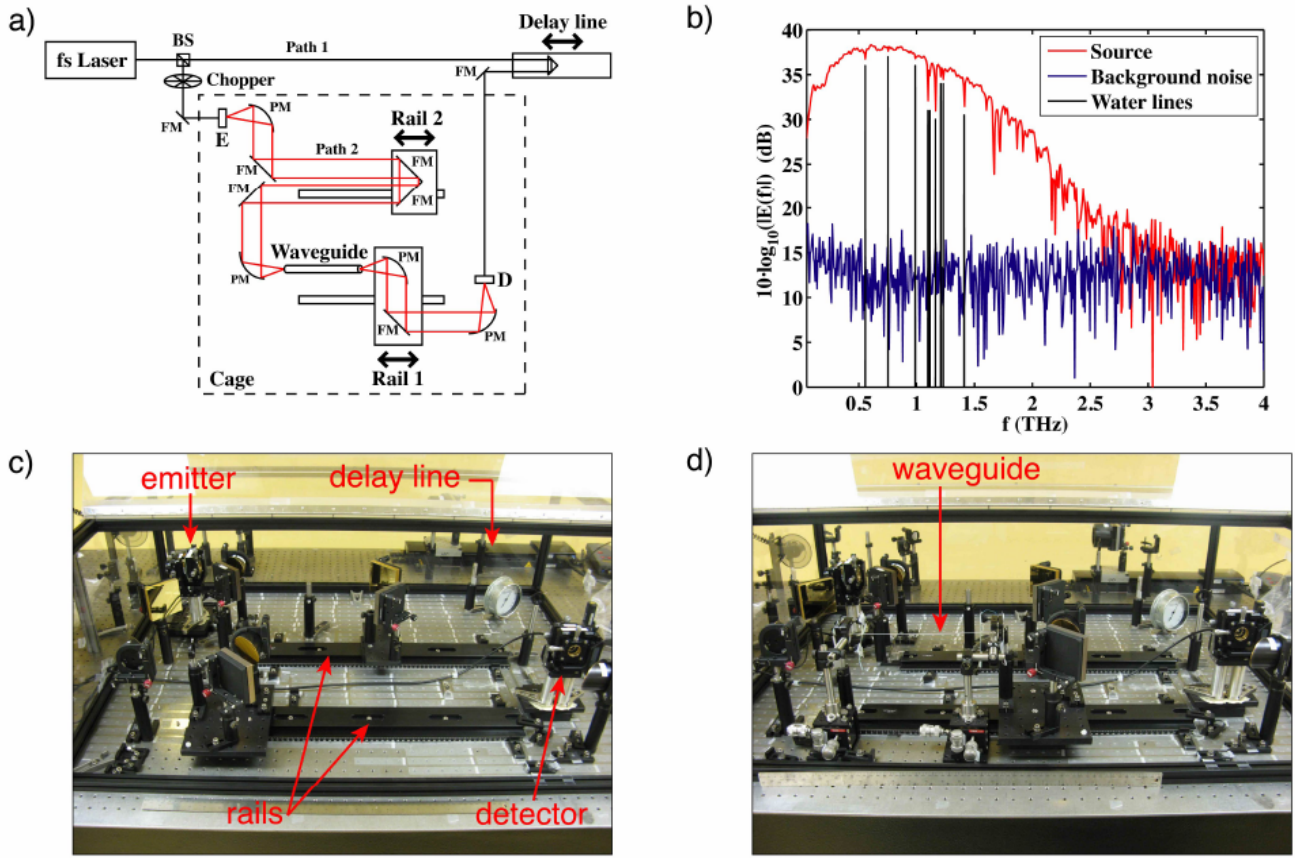

Figure 1. Tunable THz-TDS setup for waveguide transmission measurements. a) Schematic of a setup. E:Emitter, D:Detector, PM:Parabolic Mirror, BS:Beam Splitter, FM: Flat Mirror, b) Source spectrum (red) and background noise level (blue). There are traces of water vapor (black) despite efforts to purge with a nitrogen atmosphere. c), d) Photographs of a setup for different positions of the mirror assembly that allows to either perform measurements of a point sample c) or to accommodate a waveguide up to 50 cm in length d).

## 3. POROUS SUBWAVELENGTH FIBERS

### 3.1 Principles of operation

The simplest subwavelength dielectric fiber consists of a solid dielectric wire acting as a fiber core, surrounded by the low refractive index low loss cladding such as dry air (see Figure 1.a). At wavelengths larger than the fiber diameter, this simple fiber with a step-index refractive index profile supports modes that extend into the air cladding over distances that can be many times larger than the fiber diameter. It is due to significant presence of a mode in the low loss cladding that the effective modal loss due to material absorption can be made very small. As an improvement to this design, Nagel et al. proposed adding a subwavelength hole to a THz subwavelength fiber.[18] As can be seen in Figure 1.b), in this design the fraction of power guided in the air increases because of the higher field concentration within the hole. This is due to the continuity of the electric flux density normal to the dielectric interfaces. Particularly, electric field at the dielectric/air interface shows a discontinuous jump to higher value inside of the air hole because of the hole much lower refractive index compared to that of a fiber material.[19] As a further improvement, adding more holes within the subwavelength fiber (Figure 1.c) was recently proposed by our group as a mean of further reducing the fiber absorption loss by forcing a greater portion of light to guide in the dry low-loss gas filling the porous fiber core.[15, 16] Furthermore, theoretical simulations predicted not only lower absorption losses in porous fibers, but also much higher confinement to the porous fiber core and consequently lower bending losses when compared to the non-porous fibers.[17]

## 3.2 Fiber fabrication

Several fiber fabrication techniques were explored in order to fabricate porous subwavelength fibers. Main challenge for the fabrication of such fibers remains preserving highly porous thin-walled structure during fiber drawing process. All the samples presented in this paper were made from the low-density polyethylene (PE) preforms. Non-porous fibers were fabricated from the same PE material to serve as a comparison for the optical properties of porous fibers. Non-porous fiber preforms were made by fusing PE granules within a tube to form a solid cylinder that was subsequently drawn into a fiber.

### 3.2.1 Sacrificial polymer technique

The first method that was developed for porous fiber fabrication is a Sacrificial Polymer Technique, and has been described in detail earlier.[12] Schematic of the fabrication steps is presented in Figure 2(d). Within this method the rods of a sacrificial polymer (polymethyl-methacrylate (PMMA) in our case) are arranged in a hexagonal pattern without touching. Then the rest of a perform is filled with PE granules. Sacrificial material is chosen to have a significantly higher glass transition temperature. This allows melting the low viscosity PE polymer granules that consequently fill out the space between the sacrificial rods. Thus fabricated perform was then drawn into a fiber at $210^oC$. Finally, the air holes in a fiber are revealed by dissolving the sacrificial PMMA rods in tetrahydrafuran (THF) without affecting the rest of the fiber. The final fiber retains the same geometry as the preform, but with air holes in place of sacrificial rods. The presence of PMMA prevents collapse of the holes that otherwise would have happened if they were left empty. The main advantage of a sacrificial polymer technique is that drawing of porous structures is greatly facilitated as hole collapse is prevented completely during fabrication. The main disadvantage of this method is that a postprocessing step of removing the sacrificial polymer is required. Although we present a fiber with 35% porosity by area, it should be noted that this technique is very versatile and much higher porosities can easily be achieved by incorporating more sacrificial polymer into the preform (larger PMMA rods, for instance).

### 3.2.2 Microstructured molding technique

Another fabrication method that was developed for the fabrication of porous (and, in general microstructured fibers) uses casting of a fiber preform in the microstructured mold. The resultant preform features air holes which have to be pressurized during drawing to prevent hole collapse. Schematic of the fabrication steps is presented in Figure 2(e). A porous cylindrical preform, featuring 7 holes running its entire length, was fabricated by melting PE granules and solidifying the melt within a microstructured glass mold. Microstructured mold was prepared by first aligning thin-walled quartz capillaries in a hexagonal arrangement using special alignment rigs, followed by fusion of such aligned capillaries to the bottom of a large diameter quartz ampoule. The ampoule was left partially empty to allow placement of the polymer granules on top of a microstructured mold. The ampoule was then filled with granules, sealed and placed into the furnace. After melting the polymer, the polymer melt was transferred by gravity into the microstructured mold region. Upon cooling, most of the glass mold could be removed by simply pulling it from the polymer preform. Any residual glass was dissolved in the hydrofluoric acid, which had no effect on the PE perform. Porous fiber was subsequently drawn while pressurizing the preform holes. In addition to preventing hole collapse during drawing, a sufficiently large air pressure could inflate the holes and greatly increase the fiber porosity. Using molding technique followed by hole pressurization allowed us to produce fibers of very higher porosity (as high as 89%).

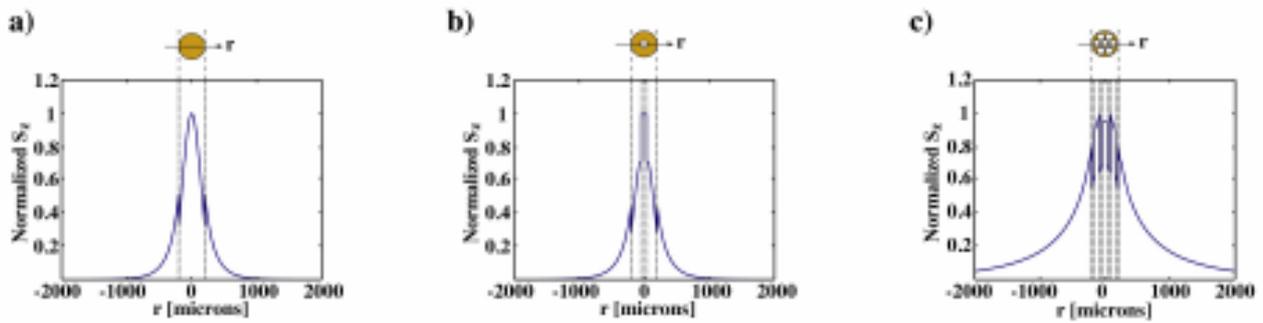

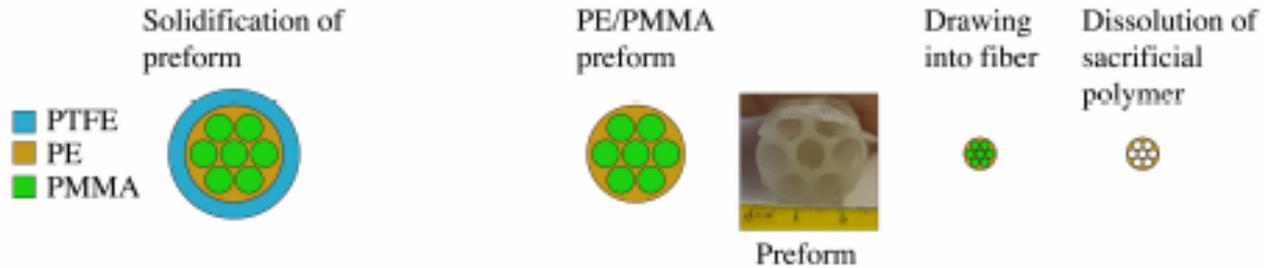

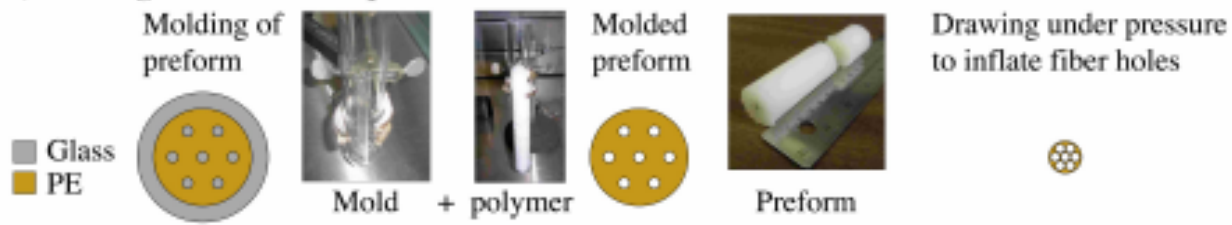

Figure 2. Schematics of various subwavelength fibers and their fabrication techniques. a)-c) Poynting vector distributions across fiber crossections for subwavelength fibers featuring 0, 1, and 7 holes, respectively. The outer diameter of all the fibers is 400μm, the diameter of all the holes is 100μm, and the frequency is 0.3 THz ($\lambda$ = 1000mm); Schematics of the d) sacrificial polymer technique and e) microstructured molding technique for the fabricating of porous subwavelength fibers.

## 4. TRANSMISSION AND LOSS MEASUREMENTS

The transmission spectra of the subwavelength fibers were measured using the THz-TDS setup described in section 2. The cutback method was used in order to evaluate modal propagation loss. During measurements the fibers were held straight by knotted threads[12] and the fiber ends were aligned with respect to the pre-installed aligning apertures using 3-axis mounts. These apertures marked the locations of focal points of the input and output off-axis parabolic mirrors. Proper alignment with respect to these apertures insured consistent coupling into and out of the fibers. Figures 3(a) and (e) present cross-sections of the measured subwavelength fibers. Porous fiber made using sacrificial polymer technique (designated as PE/PMMA fiber in Figure 3(a)) features 35% porosity, it has an outer diameter of d=450μm, hole size $d_h$=100mm, and a hole diameter to the hole-to-hole pitch ratio $d_h/L$=0.60. In turn, porous fiber made with a microstructured mold casting technique (designated as molded PE fiber in Figure 3(b)) features an outer diameter d=775mm and 86% porosity (equivalent to the hole diameter to the pitch ratio $d_h/L$=0.93). To our knowledge, currently this is the highest reported porosity for a subwavelength fiber. Diameters of the corresponding non-porous fibers are 445μm and 695μm, respectively. In Figures 3(b,f) we present the measured amplitude transmission spectra of the porous and non-porous fiber segments, up to 38.5cm in length. The lengths of the corresponding fiber segments are indicated in the figure captions. In Figures 3(c,g) we present the power propagation loss of the subwavelengths fibers. We would like to note that due to short lengths of the studied fibers it was difficult to cut them while keeping the input end fixed. Therefore, during cutback measurements we had to remove the fibers segments, cut them, and then place them back and re-align for every measurement, thus incurring small alignment errors.[20] In figures 3 (c,g) in dotted curves we also indicate error bars associated with our cut-back loss measurements. From Figures 3(c,g) we find that the propagation loss minima of the ~450μm diameter fibers are 0.024±0.004 cm$^{-1}$ at 0.190THz for the non-porous fiber, and 0.022±0.014 cm$^{-1}$ at 0.234THz for the porous fiber; for the ~700μm diameter fibers we find losses of 0.040±0.008 cm$^{-1}$ at 0.117THz for the non-porous fiber, and 0.012±0.010 cm$^{-1}$ at 0.249THz for the porous fiber. Note that transmission minima of the porous fibers are located at higher frequencies than those of the non-porous fibers of the same diameter. Moreover, spectral bandwidths of the porous fibers are larger than those for the non-porous fiber of the same diameter. Both of these observations are consistent with our prior theoretical work[15, 16]. Finally, in Figures 3(d,f) we also plot the total loss normalized with respect to the fiber segment length (-2ln(T)/L) which gives an upper bound on the value of the propagation loss < 0.02 cm$^{-1}$. Although theory predicts lower losses for the porous fibers, we are unable to distinguish between the already very small losses of the porous and non-porous fibers within the error of our experimental setup. As we demonstrate in the following section the transmission peak (bell-shaped transmission curve) results from the balance between absorption loss (at high frequencies), scattering loss (at low frequencies),[9] and frequency dependent coupling loss.

Interestingly, results shown in Figures 4(b,f) also suggests that a much larger diameter fiber of higher porosity can have similar transmission spectra to that exhibited by a smaller diameter fiber of lower porosity. This is in agreement with our prior theoretical simulations[16] where we have demonstrated that by increasing fiber porosity, fibers of diameter much larger than the wavelength of light can be designed to still guide in a single mode low-loss regime.

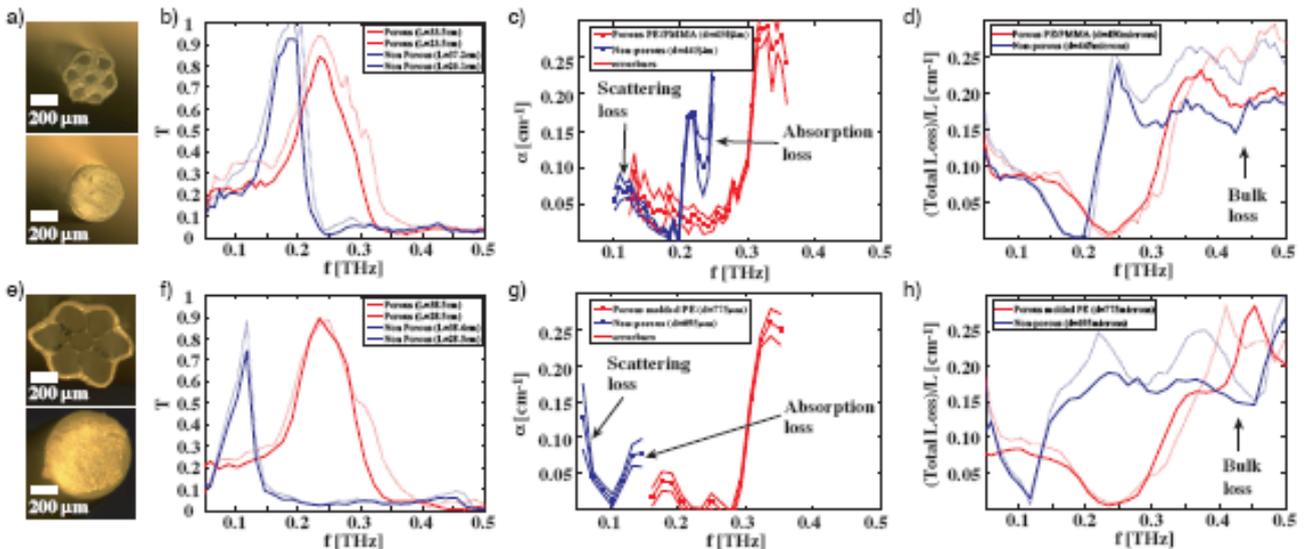

Figure 3. Transmission and loss measurements of porous and non-porous subwavelength PE fibers. Top row: small diameter fibers, Bottom row: large diameter fibers. The data for porous fibers is in red and for non-porous fibers the data is in blue. Fiber diameters and measured segment lengths are indicated in the legends. Photos a) and e) show measured fiber cross-sections; b) and f) Normalized amplitude transmission; c) and g) Propagation loss calculated from the transmission spectra using cutback technique; d) and h) Upper bound on the propagation loss from the normalized (per unit of length) total loss.

### 5. Theoretical modeling

In order to better understand transmission spectra and propagation loss results presented in Figure 3, we carried out vectorial simulations of the optical properties of the fundamental mode propagating in subwavelength fibers studied in this work. It should be noted that whereas the small diameter porous fiber (PE/PMMA fiber) retained a hexagonal arrangement of the holes, the large diameter porous fiber (molded PE fiber) had such a high porosity that it seems to be better approximated by a tube with a subwavelength-thick wall. To model the non porous fibers, an 86% porous fiber and a 35% porous fiber we therefore considered fundamental modes of a circular rod, a thin circular tube, and a 7 hole fiber. To model a non-porous rod fiber and a thin-wall-tube fiber we have used a transfer matrix code[21] to find the fundamental $HE_{11}$ mode of both fibers. In the case of a 7 hole porous fiber we have used a vectorial finite element code to find fiber modes.

In order to properly calculate the modal absorption losses the experimental value of the Polyethylen bulk absorption loss $\alpha_{PE}$ must first be. In our case, the total loss measurements of the non-porous plastic fibers (Figure 3(d)) at high frequencies suggest that the bulk loss of the LDPE polymer used in our work is $\alpha_{PE} = 0.2 cm^{-1}$.

Theoretical simulations were carried out assuming the refractive index of a PE plastic[22] to be $n_{PE} = 1.534$ with a corresponding bulk absorption loss of $\alpha_{PE} = 1 cm^{-1}$. Figure 4 presents the simulation results for each of the three geometries. The first column presents schematics of the fiber geometries. The second column presents the power absorption loss of the fundamental mode as a function of frequency. Figure 4(b) presents absorption loss of the fundamental mode of a non-porous fiber; in the figure, different curves correspond to the different fiber diameters. Figures 4(e,h) present modal losses of the porous fibers of diameters corresponding to those used in the experiment; in the figures, different curves correspond to the different values of the fiber porosity.

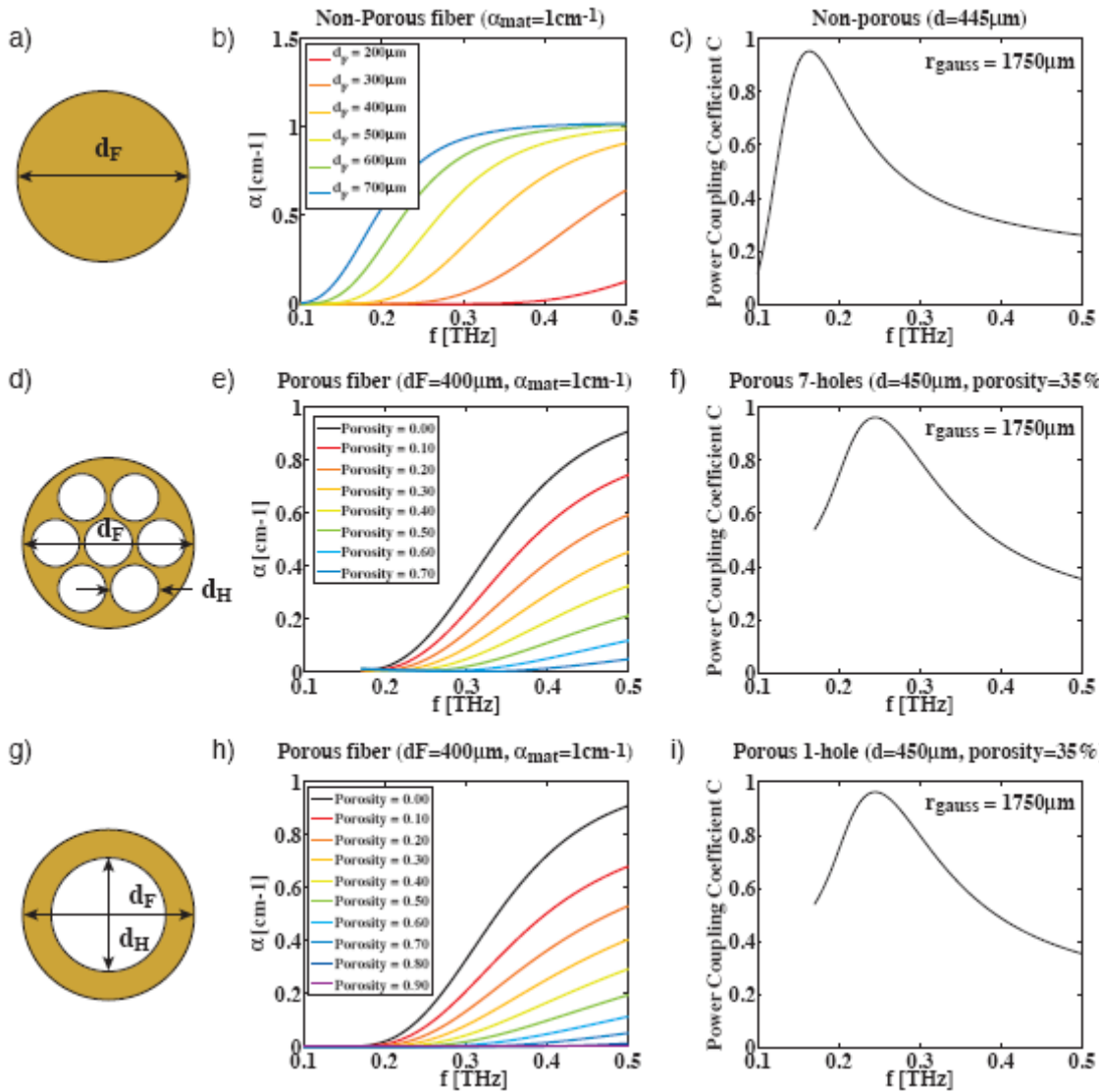

Figure 4. Modal losses of the porous and non-porous fibers as a function of the fiber geometry parameters. First row: non-porous subwavelength fiber; second row: subwavelength fiber with 7 air holes; third row: subwavelength fiber with one air hole. First column: schematics of the fiber geometries; second column: fundamental modal attenuation loss as a function of frequency; third column: coupling coefficient between a gaussian beam and the fundamental mode.

To vary fiber porosity, in both cases one simply varies the size of the air holes (porosity is defined as the ratio of the total area of the air holes to the total area of the fiber core). Behavior of the loss curves for various values of the fiber geometry parameters is easy to understand. Consider, for example, Figure 4(b). At a given frequency, choosing the fibers of smaller diameters results in a fundamental mode which is strongly delocalized beyond the lossy fiber core and into the low-loss cladding. Therefore, at a fixed frequency, fibers of smaller diameters will exhibit smaller absorption loss due to stronger modal presence in the low-loss cladding. Inversely, for a fiber of fixed diameter, when decreasing the frequency of operation, the fundamental mode dispersion relation approaches cladding light line, in other words, at lower frequencies the effective refractive index of the fundamental core mode approaches that of a cladding material. This results in a strong modal presence of the guided mode in a low-loss cladding, and as a consequence, in the reduction of the modal absorption loss at lower frequencies. Consider now Figures 4(e,h). There, one observes decrease in the fiber absorption loss for higher fiber porosities. This is easy to rationalize by noting that higher fiber porosities lead to lower refractive index contrast between the porous fiber core and a gaseous cladding. Lower refractive index contrast, in turn, means higher modal presence in the low-loss, low refractive index material (pores and cladding), hence explaining lower absorption losses of fibers featuring higher porosities.

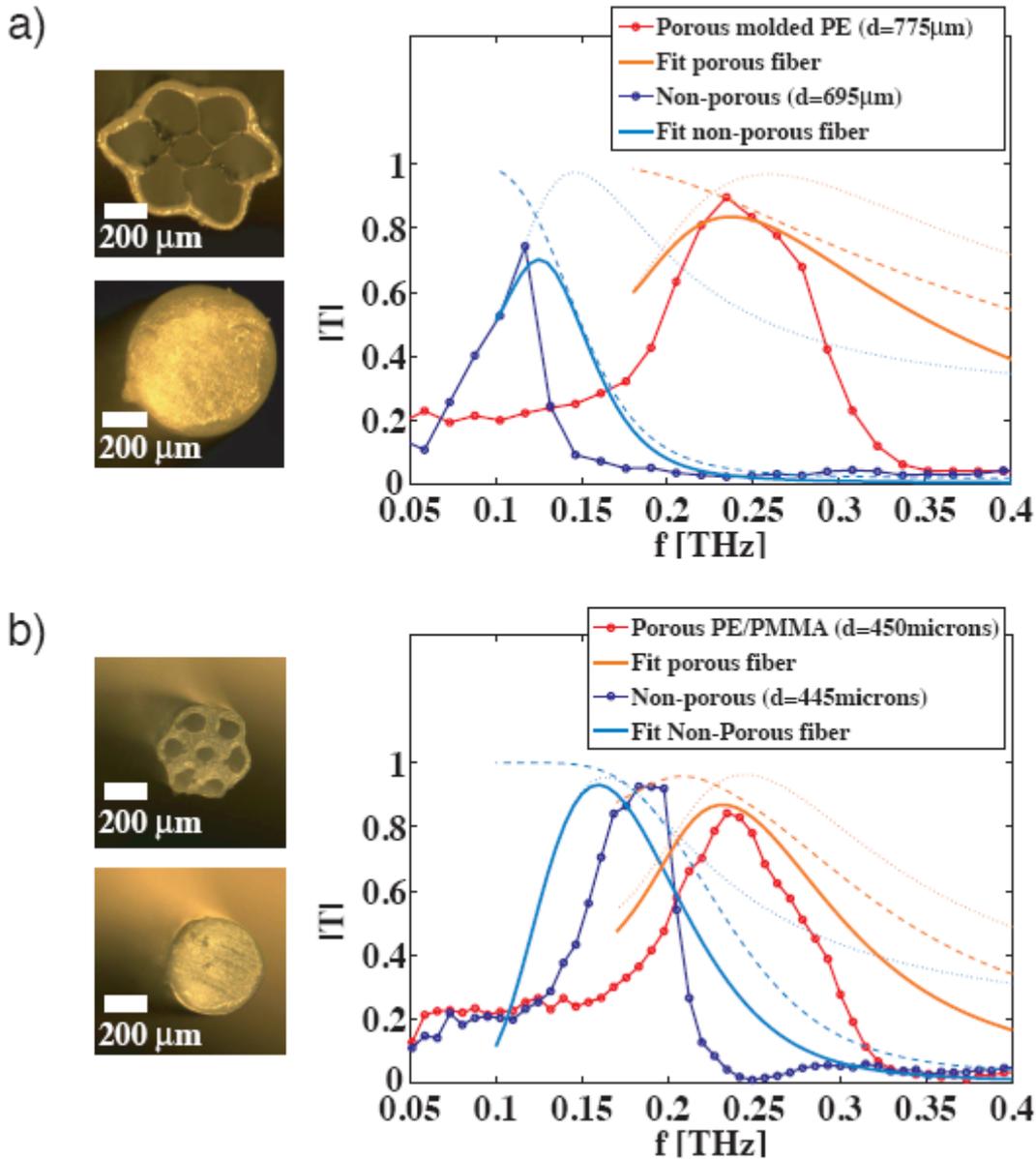

Figure 5. Theoretical fits of transmission spectra through large and small diameter subwavelength fibers. The theoretical fits (solid lines) take into account coupling loss (dotted lines) and absorption loss (dashed lines) contributions but neglect scattering losses. The calculations assumed $\alpha_{mat} = 0.2 cm^{-1}$, a porosity of 35% for the small diameter fiber, and a porosity of 72% for the large diameter fiber.

Consider now coupling losses from the gaussian-like THz beam into the fundamental modes of porous and non-porous fibers. The third column of Figure 5 presents the power coupling efficiency between linearly-polarized gaussian beam of a THz source and the fundamental mode of fibers under consideration. Using the knife-edge technique, a gaussian beam waist at the focal plane of a parabolic mirror was measured to be $2r_{guass}=3.5mm$ (beam diameter at 1/e of the field amplitude). This value was roughly independent of the operation frequency due to the use of low-dispersion parabolic mirrors in the THz-TDS setup. Note that the calculated coupling coefficients are strongly frequency dependent due to strong dependence of the modal diameters on the frequency of operation. In fact, for all the fibers there exists a frequency of optimal coupling from the gaussian beam into the fiber fundamental mode at which the beam size and the modal diameter are matched. Notably, frequency of the optimal coupling into porous fibers is always higher than that for the non-porous fibers of the same diameter, which is due to stronger modal delocalization in the non-porous fibers.

Experimentally measured transmission spectra through subwavelength fibers all feature bell-like profiles with frequencies of the transmission maxima defined by the competition between several loss mechanisms such as coupling, material absorption, and scattering on fiber imperfections[9,12]. Particularly, at higher frequencies, both the absorption loss and coupling loss increase substantially leading to a decrease in the fiber transmission (see Figures 3(c,g), and Figures 4 (b,c), for example). At lower frequencies, it is the increased coupling loss and scattering loss that lead to a decrease in the fiber transmission. Thus, at the frequency of strongest transmission, fiber mode extends sufficiently into the porous region and into the gaseous cladding to lower absorption loss, however still remaining relatively confined to the fiber core to avoid too much scattering losses on the fiber defects. As the onset of strong absorption and coupling losses for porous fibers happen at higher frequencies than those for the non-porous fibers of the same diameter (see Figure 4(b,c,e,f)), it is not surprising that experimentally measured transmission curves for the porous fibers are shifted to the higher frequencies with respect to those of the non-porous fibers (see Figures 3(c,g)).

Finally, in Figure 5 we present fits of the experimentally measured transmission curves using theoretically calculated absorption and coupling losses. The theoretical fits for the fiber transmission (solid lines) are calculated using equation (3) by taking into account coupling loss (dotted lines) and absorption loss (dashed lines) contributions (see Figure 4). Even without taking scattering loss into account we find a good correspondence between the theoretical and experimental curves. Scattering losses (not considered in this work), would further narrow the transmission peaks on the low frequency side. Although the porosity of the large diameter fiber was estimated to be 86% from the photo of a fiber cross-section, we find that a porosity of 72% yielded the best fit using the approximation of this fiber with a thin-wall tube. The greater widths of the porous fiber transmission spectra are attributed in part to the slower frequency dependence of the absorption loss curves and in part to the broader coupling loss curves. The latter being due to the slower variation of the modal size as a function of frequency for the fibers with lower refractive index contrast (porous fibers versus non-porous fibers). Finally, we would like to note that although the absorption loss of a porous fiber is predicted to be much lower than that of a non-porous fiber of the same diameter, we find experimentally that the lowest measured losses of the porous and non-porous fibers are quite similar and on the order of 0.01 cm$^{-1}$-0.02 cm$^{-1}$. We attribute this finding to the limit on the lowest obtainable loss set by the scattering loss on various fiber imperfections such as dust on a fiber surface, bulk material impurities, fiber diameter fluctuations, fluctuations in the porosity microstructure, micro- and macro-bending, etc.

## 6. Discussion and conclusions

In this work we have demonstrated low-loss THz transmission using various subwavelength Polyethylen fibers. Solid core ~40cm-long subwavelength fibers of diameters 445μm and 695μm were demonstrated to guide in the 0.14-0.21THz and in the 0.06-0.14THz spectral regions respectively with losses as low as 0.02cm$^{-1}$. Addition of porosity to the subwavelength rod-in-the-air dielectric fibers has been shown to shift the propagation loss minimum to higher frequencies and to broaden the fiber transmission spectra. Thus, porous fibers of diameters similar to those of the non-porous fibers were demonstrated to guide in the 0.16-0.31THz (fiber of 450μm diameter with 35% porosity) and in the 0.18-0.30THz (fiber of 775μm diameter with 86% porosity) spectral ranges. Interestingly, these results also suggest that a much larger diameter fiber of higher porosity can have a transmission spectrum similar to that exhibited by a smaller diameter fiber of lower porosity. We, therefore, believe that highly porous fibers can enable single mode transmission at higher frequencies beyond 0.4THz, while still exhibiting very efficient coupling to the gaussian-like several mm-diameter beam of a THz source.

In the paper we have also presented two techniques for the fabrication of highly porous fibers. Both techniques start with a fabrication of a microstructured mold which is then used to cast a THz fiber preform by melting the Polyethylen polymer and filling the mold with it. In the sacrificial polymer technique the mold is made from plastic which is drawn together with a PE polymer into a microstructured fiber; the plastic of the mold is then removed from the drawn fiber by dissolving it in the organic solvent. The main advantage of this method is that complex air microstructure can be created as hole collapse during drawing is not an issue. Main limitation of the sacrificial polymer technique is the necessity of a post-processing step to remove the plastic of a mold. Another fiber fabrication strategy presented in this paper is based on a microstructured mold casting method where the material of the mold is quartz that has to be removed before fiber drawing. The main advantage of this method is that very high precision microstructured molds can be created using standard glasswork, with an additional advantage of high purity all-quartz environment. Main disadvantage is that air microstructure has to be pressurized during drawing to prevent its collapse due to surface tension effects.

## 5. Acknowledgements
This work was sponsored in part by an NSERC Alexander Graham Bell scholarship grant. We would also like to acknowledge Canada Institute for Photonic Innovations FP3 project and Canada Research Chairs program for the financial support of this work.